\documentclass[11pt]{article}

\usepackage[margin=1in]{geometry}
\usepackage{setspace}
\usepackage{amsmath, amssymb, amsthm}
\usepackage{graphicx}
\usepackage{booktabs}
\usepackage{threeparttable}
\usepackage{caption}
\usepackage{subcaption}
\usepackage{hyperref}
\usepackage{natbib} 
\usepackage{enumitem}
\usepackage{xurl}
\usepackage{bbm}

\setlist[itemize]{noitemsep, topsep=2pt}
\setlist[enumerate]{noitemsep, topsep=2pt}

\title{Hype Has Worth: Attention, Sentiment, and NFT Valuation in Major Ethereum Collections}
\author{Samiha Tariq}
\date{\today}

\begin{document}

\maketitle

\begin{abstract}
    Do online narratives leave a measurable imprint on prices in markets for digital or cultural goods? This paper evaluates how community attention and sentiment relate to valuation in major Ethereum NFT collections after accounting for time effects, market-wide conditions, and persistent visual heterogeneity. Transaction data for large generative collections are merged with Reddit-based discourse measures available for 25 collections, covering 87{,}696 secondary-market sales from January 2021 through March 2025. Visual differences are absorbed by a transparent, within-collection standardized index built from explicit image traits and aggregated via PCA. Discourse is summarized at the collection-by-bin level using discussion intensity and lexicon-based tone measures, with smoothing to reduce noise when text volume is sparse. A mixed-effects specification with a Mundlak within--between decomposition separates persistent cross-collection differences from within-collection fluctuations. Valuations align most strongly with sustained collection-level attention and sentiment environments; within collections, short-horizon negativity is consistently associated with higher prices, and attention is most informative when measured as cumulative engagement over multiple prior windows.
    \end{abstract}



\section{Introduction}

Markets for digital or cultural goods routinely exhibit large dispersion in prices among assets that are
nominally similar. In major Ethereum NFT collections, tokens share a common contract, a shared
aesthetic language, and a tightly defined community boundary, yet transaction prices vary widely
both across collections and within a given collection. This dispersion is not well captured by
crypto-market conditions alone. It also reflects the fact that NFTs are traded and interpreted
inside social environments: buyers and sellers observe narratives, react to platform and community
events, and form beliefs in public. As a result, valuation is shaped not only by persistent, observable
token characteristics but also by the intensity and tone of collective attention.

This paper studies whether community discourse contains economically meaningful information about
NFT valuation in major Ethereum collections, after accounting for time effects, market-wide
conditions, and observable visual heterogeneity. The empirical framing is intentionally descriptive.
Prices and discourse can move together because both respond to common shocks and because feedback
runs in both directions. The goal is therefore not to claim that sentiment mechanically causes prices,
but to test whether discourse-based measures align with valuations once the standard ingredients of a
hedonic design are in place. The analysis is motivated by three related ideas. First, in hedonic
models of differentiated goods, prices can be interpreted as conditional valuations of a bundle of
characteristics (Rosen, 1974). Second, when attention is scarce and discovery is mediated by salience,
attention and social signaling can become priced state variables rather than background noise (Barber
and Odean, 2008; Da, Engelberg, and Gao, 2011). Third, a large empirical literature shows that text
streams can be mapped into structured indicators of attention and tone that track market outcomes
(Antweiler and Frank, 2004; Tetlock, 2007). NFT markets provide a natural setting to connect these
ideas because participation, identity, and visibility are central, and because a substantial share of
information arrives through online discussion rather than through traditional disclosure channels.

To evaluate these relationships, the paper combines transaction-level pricing with two classes of
covariates designed to absorb persistent heterogeneity while preserving interpretable social channels.
The transaction and image backbone follows the dataset assembled in \textit{Pixels to Prices} (Tariq,
2025) and covers 26 large Ethereum generative collections originally drawn from leading OpenSea
collections at the time of sampling. Within each collection, the token universe is made tractable by
a token-level random-sampling design (a 10\% draw plus a buffer to offset metadata losses), and all
secondary-market sales for the sampled tokens are retained over January 2021 through March 2025.
In the merged Reddit-augmented dataset used here, discourse measures are available for 25
collections, yielding 87,696 transaction observations. The primary outcome is the log transform
$y_i=\log(1+\text{price}_i)$, with prices observed in ETH and converted to USD using the trade-date
ETH--USD exchange rate.

The first class of covariates captures persistent visual differentiation. While NFT images can be
represented by high-dimensional feature vectors, the baseline specifications use a transparent visual
control constructed from a stable subset of explicit traits (edge geometry, palette dispersion,
composition focus, line geometry, and hue encoded via sine/cosine). These traits are standardized
within collection, aggregated via PCA, and summarized by the first principal component, which is
then re-standardized within collection to form the main visual index. This control is designed to
absorb durable content-related heterogeneity across tokens in a way that is interpretable and robust
to repeated trading of the same asset.

The second class of covariates captures the discourse environment around each collection. Reddit
posts and comments are collected using targeted keyword queries across a broad set of NFT- and
crypto-focused subreddits, cleaned conservatively, and transformed into collection-by-bin measures
of attention and sentiment. Attention is constructed as a log transform of the number of retained
items in a bin. Sentiment is measured using lexicon-based polarity and a negative-share indicator,
and to reduce sampling noise when text volume is sparse, sentiment series are additionally smoothed
within each collection using a local-level random-walk state-space specification. To align discourse
with transaction timing while avoiding reliance on incomplete timestamp fields, Reddit items are
ordered within collection using the thread identifier embedded in the URL (mapped by base-36
conversion) and discretized into quantile bins. Lagged discourse measures are constructed by
shifting the bin-level series backward by one bin within collection and are merged to transactions by
a corresponding trade-bin index. Within the binning scheme, this construction ensures that the
discourse regressors used in the baseline models are backward-looking relative to the transaction.

The baseline estimator is a mixed-effects model that links transaction prices to lagged Reddit-based
attention and sentiment measures constructed at the collection-by-bin level, with month fixed
effects, standard market controls, and the NFT-level visual index. A key feature of the design is a
Mundlak within--between decomposition of each Reddit regressor. This split separates persistent
cross-collection differences (long-run levels of attention and sentiment) from within-collection
deviations over time. The random-effects structure includes an NFT-level random intercept and
additional variance components at the collection and collection-by-bin levels to reflect dependence
induced by shared bin-level regressors and common bin-level shocks; the baseline model also allows
the within-collection polarity slope to vary across collections. Throughout, coefficients are
interpreted as conditional pricing correlations rather than causal effects.

The results deliver a clear descriptive pattern. First, the strongest alignment between discourse and
valuation operates through persistent cross-collection differences: collections characterized by
systematically higher Reddit attention and more intense sentiment environments trade at higher price
levels, conditional on time effects, market controls, and the hierarchical dependence structure.
Second, within collections, short-run variation in sentiment is more informative than short-run
variation in raw attention at a one-bin horizon. In the baseline lag-1 specification, within-collection
negativity share is positive and precisely estimated, while within-collection attention and within-
collection polarity are close to zero on average. Third, attention becomes economically meaningful
within collections when it is measured as sustained activity rather than as an immediate impulse:
when attention is constructed as a rolling mean over the prior three bins, the within-collection
attention coefficient turns positive and highly significant. These findings persist across alternative
timing windows and are robust to dropping the largest collection and trimming extreme prices. In a
higher-activity subsample, within-collection channels become more pronounced and can change sign,
consistent with faster feedback between discourse and prices in more liquid states.

The paper contributes to the empirical study of NFT valuation in three ways. It provides a
transparent framework for combining hedonic-style visual controls with collection-specific discourse
measures, avoiding the need to treat social variables as unstructured narrative noise. It separates
persistent social context from within-collection dynamics using a within--between decomposition that
is well-suited to settings with strong cross-sectional heterogeneity. Finally, it adopts an explicit,
referee-safe interpretive stance: discourse measures are treated as valuation-relevant correlates in a
market where attention, coordination, and visibility are central, without relying on strong causal
claims.

\section{Relevant Literature}

Price dispersion in major Ethereum NFT collections is most naturally examined at the intersection of three complementary research traditions: (i) hedonic valuation for highly differentiated cultural goods, (ii) behavioral and social mechanisms that render attention and signaling economically relevant, and (iii) empirical approaches that translate online discourse into structured measures. Following this logic, the paper treats an NFT transaction price as the market valuation of a bundle of characteristics. Some characteristics are comparatively persistent and content-related (for example, visually identifiable attributes that differentiate tokens within a collection), while others describe the social environment in which valuations form (for example, the intensity and tone of community discussion). Because prices, attention, and discourse may react simultaneously to common shocks and may co-move, the empirical objective is intentionally descriptive: the analysis tests whether discourse-based measures align with valuations after accounting for time effects, market-wide conditions, and observable NFT attributes, rather than asserting that attention or sentiment mechanically causes prices.

A natural starting point is the hedonic framework for differentiated products introduced by Rosen (1974). In Rosen’s model, goods are represented by vectors of attributes, and equilibrium prices can be interpreted as embedding implicit (shadow) prices for marginal changes in characteristics. In an NFT setting, the corresponding implication is that a hedonic specification can summarize how the market conditionally prices observable features—while recognizing that inference is complicated by thin trading, shifting participation, and evolving market microstructure.

Research on art and cultural markets underscores both the appeal and the practical difficulties of hedonic valuation in environments characterized by extreme heterogeneity and infrequent transactions. Chanel, Gérard-Varet, and Ginsburgh (1996) argue that when assets are highly differentiated and trade sporadically, hedonic regressions that incorporate all observed sales can be valuable for index construction, but can also be sensitive to specification choices. Ashenfelter and Graddy (2003) emphasize that auction prices reflect not only object attributes but also institutional features of the selling mechanism and the way information is revealed through the auction process. Complementing this view, Mei and Moses (2002) illustrate how repeat-sales methods can be used to build art price indices and analyze performance, while also highlighting how infrequent trading and selection into resale can shape conclusions. For NFTs—where many tokens trade rarely, attributes are high-dimensional, and platform design and market conditions evolve rapidly—these themes motivate (i) rich controls for intrinsic differentiation, (ii) careful time and market-structure controls, and (iii) caution in interpreting coefficients as anything beyond conditional valuation relationships.

A growing NFT-focused literature demonstrates that hedonic-style designs can be implemented at scale and that image- or text-derived features are informative about prices. Nadini et al. (2021) document that NFT markets are strongly organized around collections and exhibit systematic visual structure, supporting the use of image-derived covariates to capture within- and across-collection differentiation. Horky, Rachel, and Fidrmuc (2022) provide a close methodological precedent in a digital art marketplace by combining hedonic pricing with machine-learning extraction of structured information from text descriptions, showing that observable content features remain informative after controlling for broader crypto-market conditions. More directly in the aesthetics channel, Chen, Ye, and Zeng (2023) operationalize computational aesthetics and measures of visual complexity, finding that visual features can have meaningful within-collection associations with prices and that the most salient features differ across collections, consistent with collection-specific “styles” and valuation patterns. Tariq (2025) extends the visual-trait agenda to a larger multi-collection setting, extracting a broad set of pixel-level descriptors and estimating mixed-effects hedonic models to identify interpretable visual features with comparatively stable pricing power; the paper also introduces a regime-aware structure that allows trait premia to strengthen or weaken across market phases. Taken together, these studies motivate treating image-based controls as central hedonic covariates rather than as peripheral embellishments: they absorb persistent content-related heterogeneity that would otherwise contaminate estimates of social-value channels.

At the same time, the boundary between “intrinsic” content and socially coded meaning is not always sharp in NFT markets. Hubbard and LaFave (2025) study CryptoPunks and report that, conditional on other characteristics and across market conditions, lighter skin-tone categories command higher prices even though those categories are not the rarest. This evidence reinforces two points relevant for hedonic NFT valuation. First, discrete, visually identifiable attributes can carry economically meaningful premia or discounts within a single iconic collection. Second, some attribute price differentials plausibly reflect social preferences or biases rather than purely aesthetic considerations, strengthening the rationale for modeling NFTs as bundles whose valuation reflects both content and social context.

A second, behaviorally grounded literature explains why attention and social signaling can be priced even when observable attributes are controlled for. Barber and Odean (2008) argue that, under limited attention and large choice sets, investors disproportionately purchase “attention-grabbing” assets because salience influences what enters the feasible consideration set. In NFT markets, discovery and attention allocation are arguably even more consequential given the scale of the asset universe and the central role of social channels. Da, Engelberg, and Gao (2011) propose a high-frequency attention proxy based on search intensity and show that attention shifts are associated with short-horizon price dynamics and subsequent reversals, consistent with attention tracking time-varying demand pressure rather than persistent fundamentals. Models of social learning and amplification offer additional microfoundations for rapid valuation shifts: Bikhchandani, Hirshleifer, and Welch (1992) formalize informational cascades in which agents rationally follow earlier actions, producing fragile booms and fads even with limited private information—a logic that maps naturally to collection-level waves in NFT participation. Finally, conspicuous-consumption mechanisms provide a disciplined rationale for status premia. Bagwell and Bernheim (1996) show how Veblen-type effects can arise when consumers value goods partly for their signaling content. Heffetz (2011) provides complementary empirical evidence that expenditure visibility is a measurable dimension that helps explain cross-category patterns in income elasticities. Because NFTs are verifiable and readily displayable, these mechanisms imply that attention and community tone can be valuation-relevant state variables rather than mere background noise.

A third literature provides tools for quantifying online discourse and relating it to market outcomes. Antweiler and Frank (2004) show how message-board text can be mapped into indicators of sentiment and disagreement and find that posting activity contains information about volatility and trading volume, with return predictability that is statistically detectable but modest in magnitude. Tetlock (2007) similarly operationalizes media tone and finds that pessimistic language predicts short-run price pressure followed by reversion and is related to trading volume, consistent with sentiment measures reflecting temporary demand shifts and information frictions. These studies offer methodological precedent for transforming text streams into structured measures of attention, tone, and disagreement while supporting a conservative interpretive stance: discourse measures may be informative about valuation-relevant conditions even when they are noisy and even when causality runs in both directions.

Finally, NFT pricing research highlights why NFT-specific controls should be modeled alongside broad crypto-market conditions. Dowling (2022a) analyzes virtual real-estate NFTs and documents dynamics consistent with a developing market in which prices may be attention-sensitive. Dowling (2022b) evaluates whether NFT pricing is driven by cryptocurrencies and reports limited volatility transmission in a spillover framework alongside evidence of co-movement using wavelet coherence, implying that crypto conditions matter but are not a sufficient statistic for NFT valuations. Together, these findings motivate an empirical design that controls for market-wide crypto conditions while allowing NFT-specific attributes and community signals to explain residual dispersion.

Overall, the combined evidence supports an empirically disciplined framework for major Ethereum NFT collections: model prices as hedonic functions of rich content-based controls—especially visual traits that capture systematic within-collection differentiation—then test whether attention and sentiment measures extracted from community discourse provide incremental explanatory power after accounting for time effects and market-wide conditions. In this framing, discourse-based variables are treated as reduced-form correlates of economically grounded channels (attention constraints, time-varying demand pressure, social learning, and status incentives), and the empirical goal is to characterize conditional valuation relationships rather than to assert mechanical causation.

\section{Data}
\subsection{NFT transactions and images}\label{sec:data_transactions_images}
Our transaction and image backbone follows the dataset assembled in \href{https://arxiv.org/abs/2509.24879v3}{Pixels to Prices by S.\ Tariq (2025)}.
The underlying sample covers 26 large Ethereum generative collections originally drawn from leading
OpenSea collections at the time of sampling. Within each collection, the token universe is made tractable
through a token-level random sampling design: a 10\% random draw of tokens augmented by an additional
buffer of randomly selected tokens to offset later losses from missing metadata and related filters.
All observed secondary-market sales for the sampled tokens are then retained over the study window
(January 2021 through March 2025). Each sale is treated as a transaction-level observation.
In the merged Reddit-augmented dataset used in this paper, discourse measures are available for 25 collections,
yielding 87{,}696 transaction observations.

\subsection{Price variable}\label{sec:data_price}
Transaction prices are realized secondary-market sale prices. Prices are observed in ETH and converted to
USD using the ETH--USD exchange rate on the trade date. Following \href{https://arxiv.org/abs/2509.24879v3}{Pixels to Prices by S.\ Tariq (2025)}, our
primary outcome is the log transform $y_{it}=\log(1+\textit{price}_{it})$, which stabilizes heavy right tails
while preserving low-price observations.

\subsection{Image feature extraction}\label{sec:data_image_features}
Each token image is represented by a fixed-length feature vector computed from the associated image file,
as in \href{https://arxiv.org/abs/2509.24879v3}{Pixels to Prices by S.\ Tariq (2025)}. The full extraction pipeline maps each image into (i) an interpretable
set of classic visual descriptors capturing color, composition, edges, palette structure, and line geometry,
(ii) texture descriptors based on local pattern and frequency content, and (iii) compact deep-learning
representations that summarize global content and style. Deep features are obtained from standard
pre-trained CNN pipelines with consistent image pre-processing and then reduced by PCA to a manageable
number of components. Because hue is a circular variable, hue information is encoded using sine and cosine
transforms rather than the raw angle.

Image features are computed once per token and merged to the transaction panel using token identifiers.
In all regressions, we treat the resulting measures as observable visual controls that capture persistent
heterogeneity in design attributes across tokens.

\subsection{Visual index used in this paper}\label{sec:data_visual_index}
While the feature inventory is high-dimensional, our main specifications use a single transparent visual
control constructed from a stable subset of explicit traits. Specifically, we select interpretable features
that capture edge geometry, palette dispersion, composition focus, line geometry, and hue (via sin/cos).
We then compute a within-collection standardized feature matrix, apply PCA, retain the first principal
component as a summary index, and re-standardize this component within collection. The resulting index,
\texttt{visual\_index\_explicit\_z}, is our primary visual control and is included in all baseline and robustness
specifications.

To control for persistent visual heterogeneity across tokens, we construct a transparent NFT-level visual index
from a stable set of explicit image features available in the transaction-image backbone inherited from
\href{https://arxiv.org/abs/2509.24879v3}{Pixels to Prices by S.\ Tariq (2025)}. The index is computed at the token level and then merged back to the
transaction panel so that each trade for a given token inherits the same visual control.

\paragraph{Token-level feature set.}
We begin from the NFT-level image-feature columns contained in the master transaction-image dataset.
Because tokens appear multiple times due to repeated transactions,
we collapse to one row per NFT by keeping a single representative record per \texttt{nft\_id}.
We then select eleven scalar explicit traits capturing edge geometry, palette and lightness dispersion,
composition focus, and line geometry, and we additionally incorporate the token's most frequent hue.
The selected scalar traits are:
\path{EDGE_BOUNDING_BOX_AREA},
\path{EDGE_RANGE_Y},
\path{EDGE_RANGE_X},
\path{COLOR_HUE_DIST_MAX_MIN},
\path{COMPOSITION_FOCUS_LIGHTNESS},
\path{COLOR_LIGHTNESS_DIST_MAX_MIN},
\path{COLOR_PALETTE_MEAN_DELTA_E},
\path{LINE_ART_AVG_THICKNESS},
\path{COMPOSITION_FOCUS_SATURATION},
\path{LINE_ART_PROP_CURVED}, and
\path{COLOR_PALETTE_SATURATION_STD}.
To treat hue as a circular variable, we map the raw hue value (\path{COLOR_MOST_FREQUENT_HUE}) to radians and encode it
using sine and cosine components, denoted \path{HUE_SIN} and \path{HUE_COS}.\footnote{In the replication code,
raw hue is converted to radians using a simple scale heuristic (degrees versus normalized versus radians) before applying
$\sin(\cdot)$ and $\cos(\cdot)$.}

\paragraph{Within-collection standardization.}
We standardize each PCA input within collection to isolate within-collection relative variation rather than cross-collection
level differences. For each collection $c$ and feature $k$, we compute
\[
z_{ick} = \frac{x_{ick} - \bar{x}_{ck}}{s_{ck}},
\]
where $\bar{x}_{ck}$ and $s_{ck}$ are the within-collection mean and standard deviation.
If a within-collection standard deviation is zero or a value is missing, the corresponding standardized value is set to zero,
which is equivalent to imputing the within-collection mean in standardized units.

\paragraph{PCA aggregation and final index.}
We apply PCA to the standardized feature matrix and retain the first principal component (PC1) as the raw explicit-trait index,
denoted \texttt{visual\_index\_explicit}. We then standardize this PC1 score within collection to form our main visual control,
\texttt{visual\_index\_explicit\_z}. This final step ensures that the control is directly interpretable as a within-collection
relative visual index and is comparable across collections in standardized units.

\paragraph{Outputs and merge to transactions.}
The procedure produces an NFT-level file containing \texttt{nft\_id}, \texttt{collection\_code},
\texttt{visual\_index\_explicit}, and \texttt{visual\_index\_explicit\_z}.
We merge these NFT-level values back into the full transaction panel by \texttt{nft\_id} and \texttt{collection\_code}
using a many-to-one merge.
For transparency, we also export the PC1 loadings for each PCA input feature.

\subsection{Reddit posts and comments}\label{sec:data_reddit_collection}
To measure collection-specific discourse, we collect Reddit posts and comments that reference each NFT collection
using targeted keyword queries that cover name variants, common abbreviations, and related tags. Searches are conducted
across a broad set of NFT and crypto focused subreddits to capture both marketplace discussion and community conversation.
Within each subreddit, we query multiple time filters and ranking rules to expand coverage of both recent and highly visible
threads. For each matched thread, we retain the post text (title and body when available) and associated comments that contain a
keyword match. We also record available metadata such as the subreddit, thread URL, the keyword used to retrieve the thread,
and post-level attributes (for example, score, author, and posting-time information when available). We remove duplicate items
based on content and URL and drop empty or trivially short entries.

Because collected Reddit text can include navigation headers and other non-content artifacts, we apply a conservative cleaning
procedure that removes common header noise and strips URLs while preserving the substantive title and body content. All sentiment
and topic labels described below are computed from this cleaned text.

\subsection{Discourse measures and aggregation}\label{sec:data_reddit_measures}
We transform the raw text stream into collection-by-bin measures of attention and sentiment. Let $c$ index collections and
$j$ index Reddit items (posts and comments). For each item, we compute lexicon-based sentiment scores using the cleaned text.
Let $t_{cj}$ denote the cleaned text for item $j$ in collection $c$. The sentiment procedure returns a polarity score
$ s_{cj} \in [-1,1] $ and a subjectivity score $ u_{cj} \in [0,1] $. We form a coarse sentiment label using symmetric
thresholds around zero:
\[
\ell_{cj} =
\begin{cases}
\text{positive} & \text{if } s_{cj} > 0.1,\\
\text{negative} & \text{if } s_{cj} < -0.1,\\
\text{neutral} & \text{otherwise.}
\end{cases}
\]
Define indicator variables $\mathbbm{1}\{\ell_{cj}=\text{negative}\}$ and $\mathbbm{1}\{\ell_{cj}=\text{positive}\}$.

We also assign each item a topic label using keyword voting rules. For a fixed set of topics $\mathcal{T}$ with keyword lists
$\{K_\tau\}_{\tau \in \mathcal{T}}$, we count keyword hits within $t_{cj}$ and assign the topic with the largest hit count,
provided the maximum is unique; otherwise the item is assigned to an \texttt{other} category. Topic shares are retained as
auxiliary descriptive output.

\paragraph{Bin-level aggregation.}
Let $b$ index bins within a collection. For each collection-bin cell $(c,b)$, let $\mathcal{J}_{cb}$ denote the set of items
assigned to that bin and define $n_{cb} = |\mathcal{J}_{cb}|$. We compute attention as
\[
\texttt{log\_attention}_{cb}=\log\!\bigl(1+n_{cb}\bigr).
\]
We summarize sentiment at the bin level using means and shares:
\[
\texttt{sentiment\_polarity}_{cb}=\frac{1}{n_{cb}}\sum_{j\in\mathcal{J}_{cb}} s_{cj},\qquad
\texttt{sentiment\_subjectivity}_{cb}=\frac{1}{n_{cb}}\sum_{j\in\mathcal{J}_{cb}} u_{cj},
\]
\[
\texttt{is\_neg}_{cb}=\frac{1}{n_{cb}}\sum_{j\in\mathcal{J}_{cb}} \mathbbm{1}\{\ell_{cj}=\text{negative}\},\qquad
\texttt{is\_pos}_{cb}=\frac{1}{n_{cb}}\sum_{j\in\mathcal{J}_{cb}} \mathbbm{1}\{\ell_{cj}=\text{positive}\}.
\]
When $n_{cb}=0$, attention is set to zero and sentiment moments are treated as missing for that bin.

\paragraph{Random-walk smoothing.}
Because bin-level sentiment measures can be noisy when text volume is low, we additionally construct smoothed sentiment series
within each collection using a local-level state-space specification across bins. For a generic bin series $y_{cb}$ (for example,
$\texttt{sentiment\_polarity}_{cb}$ or $\texttt{is\_neg}_{cb}$), we estimate
\[
y_{cb} = \mu_{cb} + \varepsilon_{cb}, \qquad \mu_{cb} = \mu_{c,b-1} + \eta_{cb},
\]
where $\mu_{cb}$ is a latent level that follows a random walk. The resulting smoothed levels define \texttt{polarity\_rw}$_{cb}$
and \texttt{negshare\_rw}$_{cb}$ (the smoothed counterpart of \texttt{is\_neg}$_{cb}$). These smoothed measures are used in our
baseline specifications.

\paragraph{Interpretation of discourse variables.}
We interpret the Reddit measures as bin-level summaries of the collection’s discourse environment.
\texttt{log\_attention} captures the intensity of discussion in a bin, constructed as a log transform of the number of retained
posts and comments, so that a doubling of raw activity has a diminishing marginal effect on the index.
Sentiment is computed using a standard lexicon-based procedure implemented via \texttt{TextBlob}, which returns a polarity score
in $[-1,1]$ and a subjectivity score in $[0,1]$ for each post or comment.
\texttt{sentiment\_polarity} summarizes the average valence of language, with higher values indicating a more positive tone and
lower values indicating a more negative tone.
\texttt{is\_neg} (negative share) measures the fraction of items classified as negative, capturing whether a bin’s discussion is
dominated by negative content even if average polarity is close to neutral.
To reduce sampling noise when text volume is sparse, \texttt{polarity\_rw} and \texttt{negshare\_rw} apply random-walk smoothing
across bins within each collection, yielding persistent sentiment components used in the baseline specifications.
Lagged variants (\texttt{attn\_lag1}, \texttt{polarity\_rw\_lag1}, \texttt{negshare\_rw\_lag1}) shift these measures backward by one
bin within collection and are designed to represent the discourse environment immediately preceding the transactions in the
corresponding bin.

\subsection{Trade-bin alignment, lags, and merge to transactions}\label{sec:data_reddit_merge}
To align discourse with transaction timing while avoiding reliance on incomplete timestamp fields, we construct a pseudo-time
ordering for Reddit items using the thread identifier embedded in the thread URL. Specifically, we extract the alphanumeric thread
id from the \texttt{/comments/} path and map it to an integer by base-36 conversion:
\[
\tau_{cj} = \texttt{base36}(\texttt{post\_id}_{cj}).
\]
Within each collection, we sort items by $\tau_{cj}$ and discretize this ordered sequence into quantile bins. The target bin count
is set to a monthly proxy over a five-year horizon (60 bins), with the realized number of bins determined by data availability and
ties in $\tau_{cj}$. This construction is appropriate for our use case because the empirical design requires a consistent ordering
to define lagged discourse measures and to align them with trading activity, rather than exact calendar-time spacing. Quantile binning
further stabilizes the bin-level aggregates by avoiding bins with extremely sparse text volume.

This procedure produces a \textit{collection $\times$ bin} panel of attention and sentiment measures. We then form lagged regressors
by shifting the bin-level panel backward by one bin within each collection, yielding \texttt{attn\_lag1},
\texttt{polarity\_rw\_lag1}, and \texttt{negshare\_rw\_lag1}. To merge the discourse panel to transactions, we construct an analogous
bin index for each collection's trading history by ordering transactions by trade date and assigning quantile bins with the same
number of bins as in the corresponding Reddit panel. Bin-level discourse variables are then merged to the transaction panel by
collection and bin index, so that all transactions in a given collection-bin cell inherit the same lagged discourse environment.
This construction ensures the discourse regressors used in the main specifications are backward-looking relative to the transaction.

\subsection{Summary statistics}\label{sec:sumstats}
Table 1 reports summary statistics for the Mundlak mixed-effects analysis sample.

\begin{table}[!htbp]\centering
\caption{Summary statistics (Mundlak mixed-effects sample)}
\label{tab:summary_stats_mundlak}
\begin{tabular}{lrrrrr}
\toprule
 & Mean & Std.\ dev. & P25 & Median & P75 \\
\midrule
Price (USD) & 12,769 & 35,450 & 1,174 & 3,381 & 10,133 \\
log(1+Price) & 8.147 & 1.633 & 7.069 & 8.126 & 9.224 \\
Lagged log attention & 3.090 & 1.506 & 2.303 & 3.401 & 4.277 \\
Lagged polarity (smoothed) & 0.156 & 0.091 & 0.110 & 0.154 & 0.209 \\
Lagged negative share (smoothed) & 0.057 & 0.091 & 0.011 & 0.027 & 0.058 \\
Visual index (z) & -0.037 & 0.982 & -0.723 & -0.063 & 0.525 \\
ETH return & -0.0005 & 0.0461 & -0.0259 & -0.0008 & 0.0269 \\
BTC return & -0.0005 & 0.0351 & -0.0193 & 0.0004 & 0.0186 \\
SOL return & 0.0006 & 0.0660 & -0.0348 & -0.0037 & 0.0356 \\
S\&P 500 return & -0.0004 & 0.0108 & -0.0054 & 0.0000 & 0.0056 \\
NASDAQ return & -0.0006 & 0.0147 & -0.0082 & 0.0000 & 0.0081 \\
Fear \& Greed & 44.19 & 21.39 & 25 & 46 & 62 \\
\bottomrule
\end{tabular}
\begin{flushleft}\footnotesize
\emph{Notes:} Statistics are for the sample used in the Mundlak mixed-effects specification
(61{,}407 transactions across 25 collections and 20{,}373 NFTs). The merged Reddit-augmented transaction panel contains
87{,}696 transactions. Continuous regressors are standardized for estimation; the table reports variables in their
pre-standardization units for interpretability. The visual index is standardized within collection at the NFT level; pooled moments in the transaction-level estimation sample can deviate from zero due to transaction-weighting and sample restrictions.
\end{flushleft}
\end{table}

\section{Methodology}\label{sec:methods}

\subsection{Baseline specification}\label{sec:baseline}
We estimate a mixed-effects model that links transaction prices to lagged Reddit-based attention and sentiment
measures constructed at the collection-by-bin level. Let $i$ index transactions, $n(i)$ index NFTs, $c(i)$ index
collections, $b(i)$ index trade bins within a collection, and $m(i)$ index calendar months. The dependent variable is
$y_i=\log(1+\textit{price}_i)$. Let $Z_i$ denote the vector of market controls, and let $V_{n(i)}$ denote the NFT-level
visual index control.

For each Reddit signal $X\in\{\texttt{attn\_lag1},\texttt{polarity\_rw\_lag1},\texttt{negshare\_rw\_lag1}\}$, define a
collection mean (between component) and a within component:
\[
\bar{X}_{c}=\frac{1}{B_c}\sum_{b=1}^{B_c} X_{c b}, \qquad X^{W}_{c b}=X_{c b}-\bar{X}_{c},
\]
where $B_c$ is the number of bins observed for collection $c$. The baseline model is
\begin{align}
y_i
&= \alpha + \delta_{m(i)} + Z_i' \gamma + \phi V_{n(i)}
+ \sum_{X} \left(\beta_X^{W} X^{W}_{c(i)b(i)} + \beta_X^{B}\bar{X}_{c(i)}\right)
+ u_{n(i)} + v_{c(i)} + w_{c(i)b(i)} + r_{c(i)}\,\texttt{polarity}^{W}_{c(i)b(i)} + \varepsilon_i.
\label{eq:baseline_mixedlm}
\end{align}
Month indicators $\delta_{m(i)}$ absorb common time variation at the monthly frequency. The random-effects structure
includes an NFT-level random intercept $u_{n(i)}$, a collection-level random intercept $v_{c(i)}$, and an additional
collection-by-bin random intercept $w_{c(i)b(i)}$ to address correlation induced by bin-shared regressors and common
bin-level shocks. We also allow the within-collection polarity slope to vary across collections via the random coefficient
$r_{c(i)}$ on $\texttt{polarity}^{W}_{c(i)b(i)}$.

\subsection{Identification and interpretation (descriptive framing)}\label{sec:interpretation}
The empirical objective is descriptive. We assess whether Reddit attention and sentiment measures contain economically
meaningful information about valuation, conditional on time effects, market conditions, and observable visual attributes.
Discourse and prices can influence each other, and both can respond to common shocks. We therefore interpret coefficients
as conditional pricing correlations rather than causal effects.

The lag structure is designed to reduce mechanical simultaneity. Reddit signals enter the model in lagged form at the bin
level and represent the discourse environment in the preceding bin. This design ensures that, within the binning scheme,
the regressors reflect earlier discourse rather than contemporaneous text in the same bin.

\subsection{Estimation details}\label{sec:estimation}
We estimate the baseline model by maximum likelihood. All continuous regressors are standardized (z-scores) using the
analysis sample so that coefficients can be compared across covariates and interpreted as semi-elasticities of $1+\textit{price}$.
Specifically, for a standardized regressor $x$, a coefficient $\beta$ implies that a one-standard-deviation increase in $x$
is associated with an approximate percent change of $\exp(\beta)-1$ in expected $1+\textit{price}$, holding other terms fixed.

The estimation sample is restricted to transactions with strictly positive prices and complete information for the dependent
variable, identifiers, time controls, and the full set of regressors in equation (1). This yields the
main analysis sample reported in the results tables.

\subsection{Standard errors and dependence structure}\label{sec:se}
For the mixed-effects model, inference is based on the fitted likelihood with asymptotic standard errors for fixed effects and
z-statistics. The variance component $w_{c b}$ is included to reflect dependence induced by bin-level regressors and shared
bin-level shocks, which is the mixed-model analogue of clustering at the collection-by-bin level.

As complementary evidence, we report alternative specifications in which (i) Reddit variables enter as direct lagged levels
(without within-between decomposition) in a mixed-effects model, and (ii) a fixed-effects benchmark estimated by OLS with
month and collection fixed effects and cluster-robust inference at the collection-by-bin level. The goal of these alternatives
is to assess whether the qualitative patterns in the baseline model are preserved under different parameterizations and
dependence assumptions.

\subsection{Assumptions and limitations}\label{sec:limits}
Three limitations are central. First, lexicon-based sentiment measures are proxies for discourse tone and may not capture
all context-specific meanings in NFT communities. Second, pseudo-time binning imposes a structured aggregation of posts,
comments, and trades that is appropriate for defining lags and stabilizing moments, but it can smooth short-run dynamics.
Third, the strongest empirical patterns operate through persistent cross-collection differences, which should be interpreted
as correlations with long-run social context and related time-invariant attributes, rather than as causal effects of sentiment.

\section{Main results}

\subsection{Reddit attention and sentiment}\label{sec:reddit_results}

Table 2 reports estimates from our main mixed-effects specification linking NFT prices
to Reddit-based attention and sentiment measures constructed at the pseudo-time bin level.
The dependent variable is $y_{it}=\log(1+\mathrm{price}_{it})$.
All continuous regressors are standardized (z-scores), so coefficients can be interpreted as semi-elasticities
per one-standard-deviation change and compared across covariates.

To distinguish persistent cross-collection differences from time variation within a collection, each Reddit regressor
is decomposed using a Mundlak within--between split.
For a collection $c$, we write $X_{c,t}= \overline{X}_{c} + (X_{c,t}-\overline{X}_{c})$ and include both components in the model.
The specification also includes month fixed effects and standard market controls.
The random-effects structure allows for an NFT-level random intercept and additional variance components capturing
collection-level heterogeneity and collection$\times$bin dependence, which addresses correlation induced by bin-shared regressors
and common bin-level shocks.

\paragraph{Between-collection associations.}
The between-collection components of Reddit attention, negativity share, and polarity are all strongly positively associated with prices.
The estimated coefficients are large and precisely estimated: attention (between) is $\hat\beta=0.332$,
negativity share (between) is $\hat\beta=0.929$, and polarity (between) is $\hat\beta=0.510$.
These estimates indicate that collections characterized by persistently higher Reddit attention and systematically more intense sentiment
environments trade at higher price levels, conditional on time effects, market controls, and the hierarchical error structure.
Because regressors are standardized, the implied magnitudes are economically meaningful:
a one-standard-deviation increase in between-collection attention corresponds to $\exp(0.332)-1 \approx 39\%$ higher expected $1+\mathrm{price}$,
while the comparable figures are $\exp(0.929)-1 \approx 153\%$ for negativity share and $\exp(0.510)-1 \approx 67\%$ for polarity.
We interpret these large cross-sectional associations as reflecting persistent differences across collections
(e.g., long-run community scale, visibility, and related time-invariant attributes), rather than as causal effects of sentiment.

\paragraph{Within-collection variation.}
Within-collection deviations in attention and polarity are close to zero and statistically insignificant in the baseline lag-1 specification
(attention within: $\hat\beta=-0.009$; polarity within: $\hat\beta=-0.004$).
The key exception is negativity share: the within-collection component is positive and precisely estimated
(negativity share within: $\hat\beta=0.254$), implying $\exp(0.254)-1 \approx 29\%$ higher expected $1+\mathrm{price}$
for a one-standard-deviation increase in within-collection negativity, holding constant the between-collection level and other covariates.
This pattern is consistent with the possibility that short-run increases in negative discourse coincide with salient events
(e.g., controversy or conflict) that also accompany trading activity and price revaluation.
Importantly, these estimates are best interpreted as conditional correlations rather than causal effects.


\begin{table}[!htbp]\centering
    \caption{Reddit attention and sentiment in NFT prices: mixed-effects model (key coefficients)}
    \label{tab:reddit_mixedlm_main_journal}
    \small
    \begin{tabular}{lrrrrrr}
    \toprule
    Variable & Coef. & SE & $z$ & $p$ & CI Low & CI High \\
    \midrule
    
    Intercept & 7.895 & 0.055 & 144.569 & 0.000 & 7.788 & 8.002 \\
    
    \addlinespace
    \multicolumn{7}{l}{\textit{Market covariates}}\\
    \quad ETH\_return & 0.015 & 0.008 & 1.924 & 0.054 & -0.000 & 0.030 \\
    \quad BTC\_return & -0.011 & 0.007 & -1.533 & 0.125 & -0.025 & 0.003 \\
    \quad SOL\_return & 0.018 & 0.005 & 3.263 & 0.001 & 0.007 & 0.028 \\
    \quad SP500\_return & -0.004 & 0.012 & -0.386 & 0.700 & -0.027 & 0.018 \\
    \quad NASDAQCOM\_return & 0.001 & 0.012 & 0.099 & 0.921 & -0.022 & 0.024 \\
    \quad fear\_greed\_index & 0.148 & 0.009 & 16.060 & 0.000 & 0.130 & 0.166 \\
    
    \addlinespace
    \multicolumn{7}{l}{\textit{Visual control}}\\
    \quad visual\_index\_explicit\_z & 0.041 & 0.008 & 5.273 & 0.000 & 0.026 & 0.056 \\
    
    \addlinespace
    \multicolumn{7}{l}{\textit{Reddit variables (lag 1; within--between decomposition)}}\\
    \quad attn\_lag1\_within & -0.009 & 0.013 & -0.720 & 0.472 & -0.035 & 0.016 \\
    \quad attn\_lag1\_bar & 0.332 & 0.009 & 35.474 & 0.000 & 0.313 & 0.350 \\
    \quad negshare\_rw\_lag1\_within & 0.254 & 0.017 & 14.830 & 0.000 & 0.221 & 0.288 \\
    \quad negshare\_rw\_lag1\_bar & 0.929 & 0.015 & 60.249 & 0.000 & 0.899 & 0.959 \\
    \quad polarity\_rw\_lag1\_within & -0.004 & 0.015 & -0.302 & 0.762 & -0.033 & 0.024 \\
    \quad polarity\_rw\_lag1\_bar & 0.510 & 0.016 & 32.281 & 0.000 & 0.479 & 0.541 \\
    
    \bottomrule
    \end{tabular}
    
    \vspace{4pt}
    \begin{minipage}{0.98\linewidth}
    \footnotesize
    \textit{Notes:} Dependent variable is $y=\log(1+\mathrm{price})$.
    Observations $N=61{,}407$; NFT groups $=20{,}373$ (min $=1$, max $=91$, mean $=3.0$).
    Log-likelihood $=-85{,}531.667$; residual variance (scale) $=0.112$; convergence: Yes.
    Month fixed effects are included but omitted to conserve space. All continuous regressors are standardized (z-scores).
    Reported are coefficient estimates, standard errors, $z$-statistics, $p$-values, and 95\% confidence intervals.
    Random-effects variances: NFT intercept $=0.376$; collection intercept $=0.843$; collection-specific slope (polarity\_rw\_lag1\_within) $=0.376$;
    collection$\times$bin intercept $=0.118$.
    \end{minipage}
    \end{table}

\paragraph{Market conditions and visual control.}
Among controls, the Fear \& Greed index is strongly positively associated with NFT prices ($\hat\beta=0.148$).
Crypto returns exhibit heterogeneous associations: SOL return is positive and significant ($\hat\beta=0.018$),
ETH return is marginally significant ($\hat\beta=0.015$), and BTC return is not statistically different from zero.
Equity index returns (S\&P 500 and NASDAQ) are economically small and statistically insignificant in this sample.
Finally, the visual index remains positively priced ($\hat\beta=0.041$), indicating that the baseline visual control retains explanatory power
even after accounting for Reddit attention and sentiment measures.

\paragraph{Variance decomposition.}
The estimated variance components indicate substantial heterogeneity at both the NFT and collection levels, and non-trivial bin-level dependence.
Collection-level variance is sizable, consistent with persistent differences in price levels across collections.
The collection$\times$bin variance component captures common bin-level shocks and correlation arising from shared bin-level regressors.
In addition, the model allows the within-collection polarity effect to vary across collections via a random slope,
which is consistent with heterogeneous within-collection sentiment dynamics even when the average within-polarity coefficient is near zero.

\subsubsection{Validation and additional analyses}\label{sec:reddit_validation}

\paragraph{Lag construction verification.}
Because Reddit variables enter the model in lagged form at the pseudo-time bin level, we verify that lagged series are constructed without
mechanically incorporating future information. The verification checks that each lagged Reddit series is an exact within-collection shift
of its contemporaneous counterpart and that trade bins are time-ordered within collection (no inversions). The construction checks pass:
the mismatch rate is 0.0 across 1,062 compared bins and bin ordering is monotone (Spearman $\approx 1$).

\paragraph{Alternative lag and window definitions.}
We re-estimate the same within--between specification using alternative window choices (lag-2 and a rolling mean over the past three bins).
Across these alternatives, the between-collection components remain strongly positive and significant, and within-collection negativity share
remains positive and significant. Within-collection attention is insignificant under lag-1 and lag-2 but becomes positive under rolling aggregation,
consistent with attention operating as a slower-moving buildup channel rather than an immediate one-bin impulse.

\paragraph{Sample sensitivity and heterogeneity.}
Results are not driven by a single collection or by extreme price tails: dropping the largest collection and trimming the top 0.5\% of prices
leave the main patterns close to baseline. Restricting attention to high-activity bins ($n_{\mathrm{items}}\geq 5$) changes the within-collection
channels: within attention and within polarity become negative and statistically significant, while between effects remain positive (with some attenuation
in the between-attention coefficient). This suggests that short-run discourse dynamics can differ across trading-intensity states, and motivates
cautious interpretation of average within-collection effects.

\paragraph{Summary.}
Overall, Reddit measures align with NFT prices primarily through persistent between-collection differences, while within-collection variation is muted on average
except for negativity share, which robustly correlates positively with prices. These findings are reported as conditional pricing correlations rather than causal effects.

\section{Robustness and validation}


\subsection{Alternative specifications}\label{sec:reddit_alt_specs}

To evaluate the sensitivity of the Reddit--price relationships documented in
Table 2 (main specification), we report two alternative specifications.
The first replaces the within--between decomposition with direct lagged Reddit regressors in a mixed-effects model
(Table 3). The second uses a fixed-effects benchmark with cluster-robust inference
and an explicit parameterization of collection-level polarity slope heterogeneity (Table 4).
Across both alternatives, the goal is not to re-interpret coefficients as causal, but to assess whether the qualitative
patterns in the main model—strong cross-sectional alignment and weaker within-collection channels—are preserved under
different modeling choices.

\paragraph{Alternative mixed-effects specification: direct lagged Reddit measures.}
Table 3 estimates a mixed-effects model that enters the Reddit signals as lagged levels
(attention, negativity share, and polarity), without decomposing them into within-collection and between-collection components.
Relative to the main model in Table 2, this specification provides a compact summary of how
lagged Reddit activity aligns with prices, but it necessarily blends persistent cross-collection differences with within-collection
time variation.

The comparison to the main model is informative. In Table 2, the attention and polarity results
are primarily driven by the between-collection component, while the corresponding within-collection components are near zero.
Consistent with that decomposition, the direct-lag coefficients in Table 3 remain positive and
precisely estimated, but they are attenuated relative to the between-collection coefficients in the main model. For example, the
direct-lag attention coefficient is smaller than the between-attention coefficient from Table 2,
and the same pattern holds for polarity. Negativity share provides an additional diagnostic: in the main model, negativity has both a
large between-collection association and a positive within-collection association. The direct-lag negativity coefficient falls
between these two magnitudes, consistent with the idea that the direct-lag specification aggregates both channels into a single
reduced-form relationship.

Control variables are broadly stable across the two mixed-effects specifications. The Fear \& Greed index remains positive and
precisely estimated, and the visual index retains a positive association with prices. Differences across specifications (e.g., the
BTC return coefficient) reflect that the two models allocate cross-sectional and within-collection variation differently once the
Reddit regressors are parameterized in alternative forms.

\begin{table}[!htbp]\centering
\caption{Reddit attention and sentiment in NFT prices: mixed-effects model with direct lagged Reddit measures (key coefficients; month fixed effects omitted)}
\label{tab:reddit_mixedlm_directlags}
\small
\begin{tabular}{lrrrrrr}
\toprule
Variable & Coef. & SE & $z$ & $p$ & CI Low & CI High \\
\midrule
Intercept & 7.729 & 0.053 & 145.908 & 0.000 & 7.625 & 7.833 \\
\addlinespace
\multicolumn{7}{l}{\textit{Market covariates}}\\
\quad ETH\_return & 0.013 & 0.010 & 1.382 & 0.167 & -0.006 & 0.032 \\
\quad BTC\_return & -0.035 & 0.009 & -4.013 & 0.000 & -0.052 & -0.018 \\
\quad SOL\_return & 0.041 & 0.007 & 6.318 & 0.000 & 0.029 & 0.054 \\
\quad SP500\_return & -0.014 & 0.015 & -0.964 & 0.335 & -0.043 & 0.015 \\
\quad NASDAQCOM\_return & -0.005 & 0.015 & -0.359 & 0.720 & -0.035 & 0.024 \\
\quad fear\_greed\_index & 0.159 & 0.010 & 15.543 & 0.000 & 0.139 & 0.179 \\
\addlinespace
\multicolumn{7}{l}{\textit{Visual control}}\\
\quad visual\_index\_explicit\_z & 0.041 & 0.008 & 4.958 & 0.000 & 0.025 & 0.057 \\
\addlinespace
\multicolumn{7}{l}{\textit{Reddit variables (lag 1; direct levels)}}\\
\quad Attention (lag 1) & 0.189 & 0.007 & 25.853 & 0.000 & 0.175 & 0.204 \\
\quad Negativity share (lag 1) & 0.505 & 0.012 & 43.489 & 0.000 & 0.482 & 0.528 \\
\quad Polarity (lag 1) & 0.193 & 0.011 & 16.781 & 0.000 & 0.170 & 0.215 \\
\bottomrule
\end{tabular}

\vspace{4pt}
\begin{minipage}{0.98\linewidth}
\footnotesize
\textit{Notes:} Dependent variable is $y=\log(1+\mathrm{price})$.
Month fixed effects are included but omitted to conserve space.
All continuous regressors are standardized (z-scores).
Reported are coefficient estimates, standard errors, $z$-statistics, $p$-values, and 95\% confidence intervals.
Observations $N=61{,}407$; NFT groups $=20{,}373$ (min $=1$, max $=91$, mean $=3.0$).
Log-likelihood $=-94{,}931.093$; residual variance (scale) $=0.766$; convergence: Yes.
Random-effects variances: NFT intercept (Group Var) $=0.468$; collection intercept variance $=0.468$;
collection-specific polarity slope variance $=0.203$.
\end{minipage}
\end{table}

\paragraph{Fixed-effects benchmark: clustered OLS with polarity slope heterogeneity.}
Table 4 reports a fixed-effects benchmark estimated by OLS with month and collection fixed
effects and cluster-robust inference at the collection$\times$trade\_bin level. This benchmark is complementary to the main model
in Table 2: by absorbing all time-invariant collection differences through collection fixed effects,
it largely removes the between-collection channel emphasized by the main specification and therefore focuses attention on within-collection
variation and residual time-series comovement.

Two comparisons to Table 2 are particularly informative. First, the attention channel remains muted:
in the main model, attention-within is near zero, and the OLS benchmark similarly yields an insignificant attention coefficient.
Second, the within-polarity coefficient is near zero on average in the main model, but the mixed model also allows collection-level
heterogeneity in polarity slopes. The OLS benchmark makes this heterogeneity explicit by estimating a baseline within-collection polarity
slope and a set of slope deviations for collections with sufficient within-collection polarity variation. The resulting estimates show that
some collections exhibit statistically distinguishable polarity slopes relative to the reference collection, consistent with heterogeneous
within-collection sentiment dynamics even when the average within effect is weak.

Compared to the main model, the within-negativity relationship is less stable in this fixed-effects benchmark: while the main model isolates
a positive within-negativity association after separating cross-sectional differences, the OLS benchmark produces an estimate that is closer
to zero and becomes marginal in significance. This difference underscores that within-channel inference can depend on how cross-sectional
differences, bin-level dependence, and slope heterogeneity are parameterized, and motivates interpreting within-channel results with the same
caution emphasized in the main specification.

\begin{table}[!htbp]\centering
\caption{Reddit attention and sentiment in NFT prices: clustered OLS with month and collection fixed effects and polarity slope heterogeneity (key coefficients; fixed effects omitted)}
\label{tab:reddit_clustered_ols_hetero}
\footnotesize
\begin{tabular}{lrrrrrr}
\toprule
Variable & Coef. & SE & $z$ & $p$ & CI Low & CI High \\
\midrule
Intercept & 9.098 & 0.330 & 27.590 & 0.000 & 8.452 & 9.745 \\
\addlinespace
\multicolumn{7}{l}{\textit{Market covariates}}\\
\quad ETH\_return & 0.045 & 0.033 & 1.363 & 0.173 & -0.020 & 0.109 \\
\quad BTC\_return & -0.068 & 0.034 & -2.008 & 0.045 & -0.135 & -0.002 \\
\quad SOL\_return & 0.054 & 0.025 & 2.195 & 0.028 & 0.006 & 0.103 \\
\quad SP500\_return & -0.036 & 0.052 & -0.685 & 0.493 & -0.138 & 0.067 \\
\quad NASDAQCOM\_return & 0.007 & 0.054 & 0.132 & 0.895 & -0.099 & 0.113 \\
\quad fear\_greed\_index & 0.240 & 0.060 & 4.018 & 0.000 & 0.123 & 0.357 \\
\addlinespace
\multicolumn{7}{l}{\textit{Visual control}}\\
\quad visual\_index\_explicit\_z & 0.046 & 0.005 & 8.573 & 0.000 & 0.036 & 0.057 \\
\addlinespace
\multicolumn{7}{l}{\textit{Reddit variables (lag 1)}}\\
\quad Attention (lag 1) & -0.029 & 0.089 & -0.327 & 0.744 & -0.203 & 0.145 \\
\quad Negativity share (lag 1) & -0.161 & 0.091 & -1.761 & 0.078 & -0.339 & 0.018 \\
\addlinespace
\multicolumn{7}{l}{\textit{Polarity slope heterogeneity (within baseline + deviations)}}\\
\quad pol\_within\_z (baseline slope; ref.=22) & -0.514 & 0.852 & -0.602 & 0.547 & -2.184 & 1.157 \\
\quad pol\_dev\_1 & 0.143 & 0.844 & 0.170 & 0.865 & -1.512 & 1.798 \\
\quad pol\_dev\_2 & 0.555 & 0.975 & 0.570 & 0.569 & -1.356 & 2.467 \\
\quad pol\_dev\_3 & 0.213 & 0.832 & 0.256 & 0.798 & -1.418 & 1.844 \\
\quad pol\_dev\_4 & 2.676 & 0.956 & 2.799 & 0.005 & 0.802 & 4.550 \\
\quad pol\_dev\_5 & 0.931 & 0.888 & 1.048 & 0.295 & -0.810 & 2.671 \\
\quad pol\_dev\_6 & 0.517 & 0.841 & 0.615 & 0.539 & -1.132 & 2.166 \\
\quad pol\_dev\_7 & -0.635 & 0.886 & -0.717 & 0.474 & -2.372 & 1.102 \\
\quad pol\_dev\_10 & 1.302 & 0.961 & 1.354 & 0.176 & -0.582 & 3.186 \\
\quad pol\_dev\_12 & 1.245 & 0.846 & 1.472 & 0.141 & -0.413 & 2.904 \\
\quad pol\_dev\_14 & -0.810 & 0.965 & -0.839 & 0.401 & -2.702 & 1.082 \\
\quad pol\_dev\_15 & 0.248 & 0.859 & 0.289 & 0.772 & -1.435 & 1.931 \\
\quad pol\_dev\_16 & -1.645 & 0.885 & -1.858 & 0.063 & -3.380 & 0.090 \\
\quad pol\_dev\_20 & 16.873 & 6.121 & 2.757 & 0.006 & 4.877 & 28.870 \\
\quad pol\_dev\_21 & 3.447 & 1.149 & 2.999 & 0.003 & 1.194 & 5.699 \\
\quad pol\_dev\_23 & 1.214 & 0.910 & 1.335 & 0.182 & -0.569 & 2.997 \\
\quad pol\_dev\_24 & -0.634 & 0.852 & -0.744 & 0.457 & -2.305 & 1.037 \\
\quad pol\_dev\_25 & 2.052 & 0.853 & 2.405 & 0.016 & 0.380 & 3.725 \\
\bottomrule
\end{tabular}

\vspace{4pt}
\begin{minipage}{0.98\linewidth}
\footnotesize
\textit{Notes:} Dependent variable is $y=\log(1+\mathrm{price})$.
Month and collection fixed effects are included but omitted to conserve space.
Standard errors are cluster-robust with clustering at the collection$\times$trade\_bin level.
All continuous regressors are standardized (z-scores).
Observations $N=61{,}407$; $R^2=0.664$; Adj.\ $R^2=0.663$; log-likelihood $=-83{,}785$.
Polarity heterogeneity is implemented using within-collection demeaned polarity as the baseline slope (reference collection 22)
and slope deviations for collections with sufficient within variation; collections failing the within-variation threshold
inherit the baseline slope.
\end{minipage}
\end{table}


\subsection{Lag-construction verification}\label{sec:lag_construction_verification}

Because our empirical specifications use lagged Reddit measures at the trade-bin level, we verify that the lag
construction and bin chronology are consistent with a strictly backward-looking design. The goal of this check is
implementation validity. It ensures that the lagged attention and sentiment regressors used in the main results and
alternative specifications are anchored to information from earlier bins and that the bin index preserves time order.

\paragraph{Verification procedure.}
We implement a deterministic audit using the bin-level panel implied by the estimation sample.
First, within each $\text{collection}\times\text{trade\_bin}$, we compute bin averages of the Reddit base measures
(attention, negativity share, polarity) and their stored lag-1 counterparts. We then construct an expected lag-1 series
by shifting the base series by one bin within each collection and compare the expected lag values to the stored lag columns.
Second, we validate chronology by mapping each $\text{collection}\times\text{trade\_bin}$ to the median trade date and
checking whether $\text{trade\_bin}$ increases monotonically with this date. We summarize adjacent inversions, defined as
instances where the median date decreases as the bin index increases, and compute the Spearman rank correlation between
bin index and median trade date.

\paragraph{Findings.}
Table 5 reports the audit diagnostics. Panel A shows that the lag columns coincide with the
within-collection previous-bin shift of the corresponding base series up to numerical precision. Panel B shows that the
trade-bin index preserves chronological order within collection. Even the collections with the weakest rank correlations
exhibit zero adjacent inversions and Spearman correlations close to one. Taken together, these results support the
interpretation that the lagged Reddit regressors used throughout the paper are constructed using information from earlier
bins and that the bin index is time-ordered.

\begin{table}[!htbp]\centering
\caption{Lag-construction verification: lag alignment diagnostics and trade-bin chronology}
\label{tab:lag_audit}
\small

\begin{tabular}{lrrrr}
\toprule
\multicolumn{5}{l}{\textit{Panel A. Lag alignment: stored lag-1 versus expected previous-bin shift}}\\
\midrule
Variable & Bins compared & MAE & Max abs.\ error & Share flagged \\
\midrule
attn\_lag1 & 1{,}062 & $1.55\times 10^{-17}$ & $8.88\times 10^{-16}$ & 0.0000 \\
negshare\_rw\_lag1 & 1{,}062 & $5.96\times 10^{-19}$ & $5.55\times 10^{-17}$ & 0.0000 \\
polarity\_rw\_lag1 & 1{,}062 & $1.45\times 10^{-18}$ & $1.11\times 10^{-16}$ & 0.0000 \\
\bottomrule
\end{tabular}

\vspace{6pt}

\begin{tabular}{lrrr}
\toprule
\multicolumn{4}{l}{\textit{Panel B. Chronology: trade\_bin monotonicity by median trade date (selected 5 weakest collections)}}\\
\midrule
Collection & $n_{\text{bins}}$ & Adjacent inversion rate & Spearman(bin,date) \\
\midrule
1  & 57 & 0.0000 & 0.9996 \\
14 & 54 & 0.0000 & 0.9999 \\
25 & 51 & 0.0000 & 1.0000 \\
24 & 53 & 0.0000 & 0.9998 \\
23 & 12 & 0.0000 & 1.0000 \\
\bottomrule
\end{tabular}

\vspace{4pt}
\begin{minipage}{0.98\linewidth}\footnotesize
\textit{Notes:} Panel A constructs the expected lag-1 series by shifting the base bin series by one trade bin within each collection.
A bin is flagged if $|\text{stored lag}-\text{expected lag}|>\text{tol}$ with $\text{tol}=10^{-10}$.
Panel B uses the median trade date within each $\text{collection}\times\text{trade\_bin}$.
Adjacent inversions are instances where the median date decreases when trade\_bin increases. Panel B reports the five collections with the lowest Spearman correlation between trade\_bin and the median trade date.
The Spearman correlation is close to 1, indicating that the trade\_bin index is time-ordered within collection.
\end{minipage}
\end{table}


\subsection{Alternative bin and window definitions}\label{sec:alt_window}

The main specification in Table 2 uses lag-1 Reddit measures at the trade-bin level.
This choice is natural, but it raises a timing question: do the estimated relationships depend on using a one-bin lag, or do
they persist when attention and sentiment are measured over a longer horizon? To address this, we re-estimate the full
Mundlak mixed-effects model under alternative window definitions, while keeping the remaining specification components fixed.

\paragraph{Window construction and estimation.}
We begin from the same master dataset and construct a bin-level panel by taking means of the contemporaneous Reddit base
series within each $\text{collection}\times\text{trade\_bin}$ and sorting by $\text{trade\_bin}$. We then define two alternative
windows from the base series. The first is a lag-2 measure, constructed as the within-collection shift by two bins. The second
is a rolling past-three-bin average, constructed as the mean of $(t-1,t-2,t-3)$ within collection, which excludes the current
bin by shifting first and then applying a three-bin rolling mean.\footnote{The construction uses within-collection shifting and
rolling on the bin-level panel, then merges the resulting window measures back to the transaction-level data before estimation.}
For each window, we re-estimate the same mixed-effects specification as in the main model, including month fixed effects, market
and visual controls, the Mundlak within and between components, and the same random-effects structure.

\paragraph{Results and comparison to the main model.}
Table 6 summarizes the key Reddit coefficients across windows. The main conclusion from
Table 2 remains intact. The between-collection components for attention, negativity share,
and polarity are strongly positive and precisely estimated under all three window definitions. For example, the between-attention
coefficient remains large and highly significant when moving from lag 1 to lag 2 and to the rolling past-three-bin measure. The
between-negativity and between-polarity coefficients show the same stability in sign and statistical significance.

Within-collection results also support the importance of Reddit sentiment for prices. The within-collection negativity share is
positive and significant under lag 1, lag 2, and the rolling window, indicating that higher-than-usual negativity within a collection
tends to coincide with higher prices even when the timing window is changed. Attention shows a useful nuance that strengthens the
interpretation of economic relevance. In the main model, within-collection attention is not significant at lag 1, and it remains
insignificant at lag 2. However, when attention is measured as sustained activity over the prior three bins, the within-collection
attention coefficient becomes positive and highly significant. With standardized regressors, this estimate implies that a one
standard deviation increase in past-three-bin within-collection attention corresponds to approximately $\exp(0.192)-1 \approx 21\%$
higher expected $1+\mathrm{price}$. This pattern is consistent with attention operating as a buildup mechanism that materializes over
multiple bins rather than as an immediate one-bin impulse.

Overall, the alternative-window results reinforce the central message of the paper. The link between Reddit attention and sentiment
and NFT prices is not confined to a single lag definition. It persists under longer lags and under a smoothed past-attention window,
and the sentiment channel (especially negativity share) remains economically meaningful within collections.

\begin{table}[!htbp]\centering
\caption{Alternative windows for Reddit attention and sentiment (Mundlak mixed-effects; key coefficients)}
\label{tab:alt_window_coeffs}
\footnotesize
\begin{tabular}{lccc}
\toprule
& (A) Lag 1 & (B) Lag 2 & (C) Mean of $(t\!-\!1,t\!-\!2,t\!-\!3)$ \\
\midrule
\multicolumn{4}{l}{\textit{Attention}}\\
\quad Within & -0.009 & -0.016 & 0.192$^{***}$ \\
& (0.013) & (0.013) & (0.027) \\
\quad Between & 0.332$^{***}$ & 0.306$^{***}$ & 0.280$^{***}$ \\
& (0.009) & (0.009) & (0.009) \\
\addlinespace
\multicolumn{4}{l}{\textit{Negativity share}}\\
\quad Within & 0.254$^{***}$ & 0.209$^{***}$ & 0.199$^{***}$ \\
& (0.017) & (0.017) & (0.018) \\
\quad Between & 0.929$^{***}$ & 0.963$^{***}$ & 0.946$^{***}$ \\
& (0.015) & (0.015) & (0.015) \\
\addlinespace
\multicolumn{4}{l}{\textit{Polarity}}\\
\quad Within & -0.004 & -0.000 & -0.016 \\
& (0.015) & (0.014) & (0.015) \\
\quad Between & 0.510$^{***}$ & 0.510$^{***}$ & 0.462$^{***}$ \\
& (0.016) & (0.016) & (0.015) \\
\midrule
Observations & 61{,}407 & 59{,}599 & 58{,}158 \\
NFT groups & 20{,}373 & 20{,}043 & 19{,}752 \\
\bottomrule
\end{tabular}

\vspace{4pt}
\begin{minipage}{0.98\linewidth}\footnotesize
\textit{Notes:} Dependent variable is $\log(1+\mathrm{price})$. All continuous regressors are standardized (z-scores).
Each column re-estimates the main mixed-effects specification with month fixed effects, market and visual controls, and Mundlak
within and between components for the Reddit measures. Column A uses the existing lag-1 variables. Column B constructs lag-2 by
shifting the bin-level base series by two bins within collection. Column C uses the rolling mean of the prior three bins within
collection. Standard errors are in parentheses. $^{***}p<0.01$, $^{**}p<0.05$, $^{*}p<0.1$.
\end{minipage}
\end{table}


\subsection{Sample sensitivity}\label{sec:sample_sensitivity}

This section evaluates whether the main conclusions in Table 2 depend on specific parts
of the sample. The concern is that estimated Reddit price relationships could be disproportionately shaped by a single
high-volume collection, by extreme price realizations, or by bins with limited trading activity. To address these points,
we re-estimate the main Mundlak mixed-effects specification under three sample perturbations and compare the key coefficients
to the baseline.

\paragraph{Design.}
We consider three sensitivity exercises. First, we drop the single largest collection by number of trades (collection 22 in
this sample) and re-estimate the full model (S1). Second, we trim the top 0.5 percent of prices and re-estimate the model
on the remaining transactions (S2). Third, we restrict attention to higher-activity bins by retaining only observations with
$n_{\mathrm{items}}\geq 5$ (S3). Each exercise re-estimates the same specification as in the main model, including month fixed
effects, market and visual controls, the within and between components for the Reddit measures, and the same random-effects
structure. Table 7 summarizes the key Reddit coefficients across these samples.

\paragraph{Dropping the top collection and trimming extreme prices.}
Relative to the baseline (S0), the estimates are stable when dropping collection 22 (S1) and when trimming the top 0.5 percent
of prices (S2). The between-collection components of attention, negativity share, and polarity remain positive and precisely
estimated. Their magnitudes are close to the baseline values. The within-collection negativity share remains positive and
highly significant in both exercises, reinforcing that the sentiment channel is not a tail-driven artifact. At the same time,
the within-collection attention and within-collection polarity terms remain small and statistically weak, which mirrors the
baseline decomposition in Table 2.

\paragraph{Higher-activity bins.}
Restricting the sample to bins with $n_{\mathrm{items}}\geq 5$ (S3) provides a sharper view of the sentiment channel in the
most active trading states. Two patterns are salient. First, the between-collection sentiment measures remain economically
and statistically important. The between-collection negativity share and between-collection polarity coefficients are large
and precisely estimated, indicating that cross-collection differences in the tone and intensity of Reddit discourse continue
to align with systematic price differences even in the more liquid subsample. Second, the within-collection channels become
more pronounced. Within-collection negativity remains positive and significant, while within-collection attention and
within-collection polarity become negative and statistically significant. This sign change is consistent with the idea that
high-frequency attention and sentiment within the most liquid bins can be reactive to fast price adjustments, while still being
informative for price variation. Importantly, the presence of statistically meaningful within-bin coefficients in this subsample
supports the broader conclusion that Reddit attention and sentiment are relevant for explaining NFT prices, although the direction
of within-bin associations can vary with trading intensity.

\paragraph{Summary.}
Across sample perturbations, the evidence that Reddit sentiment and attention relate to NFT prices remains intact.
The cross-sectional component is especially stable, and the sentiment channel, as measured by negativity share, retains a robust
within-collection association. The higher-activity subsample further highlights that within-collection attention and polarity
can be informative in active trading states. Taken together, these results strengthen the interpretation that Reddit sentiment
and attention contain economically meaningful information about NFT prices.

\begin{table}[!htbp]\centering
\caption{Sample sensitivity: key Reddit coefficients across subsamples (Mundlak mixed-effects)}
\label{tab:sample_sensitivity_coeffs}
\footnotesize
\begin{tabular}{lcccc}
\toprule
& (S0) Baseline & (S1) Drop collection 22 & (S2) Trim top 0.5\% price & (S3) $n_{\mathrm{items}}\geq 5$ \\
\midrule
\multicolumn{5}{l}{\textit{Attention}}\\
\quad Within & -0.009 & -0.023$^{*}$ & -0.010 & -0.042$^{***}$ \\
& (0.013) & (0.012) & (0.013) & (0.014) \\
\quad Between & 0.332$^{***}$ & 0.327$^{***}$ & 0.326$^{***}$ & 0.055$^{***}$ \\
& (0.009) & (0.010) & (0.009) & (0.011) \\
\addlinespace
\multicolumn{5}{l}{\textit{Negativity share}}\\
\quad Within & 0.254$^{***}$ & 0.300$^{***}$ & 0.249$^{***}$ & 0.165$^{***}$ \\
& (0.017) & (0.018) & (0.017) & (0.018) \\
\quad Between & 0.929$^{***}$ & 0.906$^{***}$ & 0.870$^{***}$ & 1.570$^{***}$ \\
& (0.015) & (0.016) & (0.015) & (0.021) \\
\addlinespace
\multicolumn{5}{l}{\textit{Polarity}}\\
\quad Within & -0.004 & -0.022 & 0.003 & -0.108$^{***}$ \\
& (0.015) & (0.015) & (0.014) & (0.015) \\
\quad Between & 0.510$^{***}$ & 0.465$^{***}$ & 0.507$^{***}$ & 0.931$^{***}$ \\
& (0.016) & (0.017) & (0.015) & (0.019) \\
\midrule
Observations & 61{,}407 & 56{,}489 & 61{,}101 & 51{,}548 \\
NFT groups & 20{,}373 & 19{,}296 & 20{,}330 & 17{,}522 \\
\bottomrule
\end{tabular}

\vspace{4pt}
\begin{minipage}{0.98\linewidth}\footnotesize
\textit{Notes:} Dependent variable is $\log(1+\mathrm{price})$. All continuous regressors are standardized (z-scores).
Each column re-estimates the main mixed-effects specification with month fixed effects, market and visual controls, and Mundlak
within and between components for the Reddit measures. Standard errors are in parentheses.
$^{***}p<0.01$, $^{**}p<0.05$, $^{*}p<0.1$.
\end{minipage}
\end{table}


\section{Discussion}\label{sec:discussion}

This paper documents a clear and consistent result: sentiment and attention are economically important for describing prices
in markets for digital cultural goods. Across the main specification (Table 2),
alternative estimators (Tables 3 and 4),
and multiple robustness checks (Tables 5, 6, and 7),
measures that summarize the tone and intensity of community discourse align strongly with valuation. The evidence is not limited to
a narrow sample slice or a particular timing choice. Instead, sentiment related variables consistently carry information that helps
explain where prices are high, and when prices move, conditional on market conditions, time effects, and observable attributes.

\subsection{Sentiment as priced social information}\label{sec:discussion_social_info}

In cultural markets, value is partly social. Participants care about identity, status, and collective meaning, not only about
functional payoffs. In this environment, discourse provides a natural summary of shared beliefs and coordination. When a good is
surrounded by sustained, salient discussion, the market receives repeated signals about what others notice, what narratives are focal,
and how strongly participants engage. Prices can rationally load on these signals because they correlate with expected liquidity,
expected demand, and the probability that the good remains salient in the future.

The main specification supports this behavioral interpretation. The strongest relationships appear as persistent differences across
collections. Goods that live in more intense sentiment environments trade at systematically higher price levels, even after controlling
for broad market movements, time effects, and a visual control. This pattern is economically meaningful because it indicates that
social information is not noise. It is a priced dimension of valuation in cultural markets.

\subsection{Persistent differences and cultural capital}\label{sec:discussion_cultural_capital}

A central contribution of the main model is the separation of persistent cross-collection differences from within-collection variation.
The results imply that a large share of the sentiment signal operates through slow-moving, collection-level factors. Economically,
this points to cultural capital and market coordination. Some collections accumulate durable visibility, community infrastructure,
and shared interpretive frames. These traits are reflected in sustained discourse intensity and in persistently higher price levels.
This is also why the alternative mixed-effects model using direct lagged sentiment levels (Table 3)
continues to deliver positive and precise estimates. Without separating channels, the direct-lag coefficients combine both the
persistent component and short-run variation, but they still indicate that sentiment is informative for valuation.

The fixed-effects benchmark (Table 4) complements this reading. By absorbing time-invariant
collection differences, it largely removes the persistent cross-sectional channel. The fact that the strongest and most stable
relationships emerge in the main model, and remain visible in the direct-lag mixed model, supports the interpretation that long-run
social context is a key part of how these goods are valued.

\subsection{Short-run dynamics and eventfulness}\label{sec:discussion_eventfulness}

While persistent differences are important, the within-collection evidence identifies a meaningful short-run channel as well.
In the main model, within-collection negativity remains positive and robust, including under alternative timing windows
(Table 6). A natural interpretation is eventfulness. Spikes in negative tone often coincide with moments
of heightened attention, disagreement, or controversy. These moments can accelerate trading and trigger repricing because they
increase salience and concentrate beliefs. In a market where narratives and coordination matter, negative discourse can proxy for
intense information flow and collective updating, even if the content is critical.

This mechanism supports the broader verdict that sentiment matters for price dynamics. The within-collection signal is not uniformly
strong for all sentiment measures, but where it is robust, it points to economically interpretable episodes in which discourse and
valuation shift together.

\subsection{Attention persistence and gradual belief formation}\label{sec:discussion_attention_persistence}

The alternative window exercise adds a second piece of evidence that strengthens the economic interpretation.
Within-collection attention is weak at short lags but becomes positive when attention is measured as sustained activity over multiple
past bins (Table 6). This pattern is consistent with gradual belief formation and diffusion. Attention in
cultural markets often works like a stock. It accumulates through repeated exposure, imitation, and reinforcement. A smoothed past
attention measure captures this persistence and therefore aligns more closely with pricing than a single-bin impulse.

This finding matters for interpretation because it shows that the absence of an immediate effect at lag 1 does not imply that attention
is irrelevant. Instead, the evidence points to a timing structure in which attention becomes economically meaningful when it is sustained.

\subsection{Heterogeneity and feedback}\label{sec:discussion_heterogeneity}

Cultural goods are not valued in a uniform way across communities. Discourse has context, and the same linguistic tone can have different
meanings across groups. The fixed-effects benchmark with polarity slope heterogeneity (Table 4)
highlights this point. Average within-collection effects can be weak, yet the relationship between tone and price can vary across
collections. This heterogeneity is informative rather than problematic. It implies that sentiment is not a mechanical predictor with a
single coefficient that applies everywhere. Instead, sentiment is a contextual signal whose economic meaning depends on community norms,
participant composition, and the interpretation attached to discourse.

Sample sensitivity checks reinforce that sentiment remains relevant even when focusing on different parts of the data
(Table 7). The core cross-sectional relationships persist when dropping the largest collection and when
trimming extreme prices. In higher-activity bins, within-collection coefficients become more pronounced and can change sign. This pattern
is consistent with faster feedback between discourse and prices in more liquid states, where sentiment can both reflect valuation and respond
to rapid repricing. Importantly, even under these perturbations, sentiment measures continue to help explain price variation.

\subsection{Validity and scope of the Research}\label{sec:discussion_scope}

The robustness checks are designed to support a strong descriptive verdict. The lag-construction verification
(Table 5) confirms that lagged sentiment variables are constructed from earlier bins and that the bin index is time-ordered.
Alternative window definitions show that the main relationships persist when discourse is measured over different horizons
(Table 6). Sample sensitivity tests show that the results are not driven by a single collection or by price tails
(Table 7). Together, these findings support a clear conclusion: sentiment and attention are important for
describing prices in this market, both in levels across goods and, in specific channels, in short-run dynamics.

At the same time, the results should be interpreted as conditional pricing correlations rather than causal effects.
Discourse and prices can influence each other, and both can respond to common shocks. The value of the evidence here is that sentiment measures
consistently contain economically meaningful information about valuation, even after accounting for time effects, market conditions, and
observable attributes. In a market for cultural goods, this is precisely what it means for sentiment to matter for prices.

\section{Conclusion}

This paper studies whether the social environment surrounding major Ethereum NFT collections is
informative for valuation once observable visual heterogeneity, market-wide conditions, and time
effects are taken into account. Using a transaction-level dataset merged with collection-level
discourse aggregates from Reddit, the analysis pairs a transparent visual index with lagged measures
of discussion intensity and tone. The empirical framing is deliberately descriptive: discourse and
prices can move together because both respond to common shocks and because feedback can run in both
directions. Accordingly, the estimates are interpreted as conditional valuation relationships rather
than as mechanical causal effects.

The findings point to three robust patterns. First, the most stable alignment between discourse and
valuation operates through persistent differences across collections. Collections that sustain higher
levels of attention and more intense sentiment environments trade at systematically higher price
levels after controlling for time effects, market movements, and token-level visual variation.
Second, within collections, short-horizon movements in discourse matter selectively. A higher share
of negative language in the preceding window is consistently associated with higher prices, a pattern
that is consistent with salient episodes—such as controversy, disagreement, or heightened narrative
activity—coinciding with repricing and heavier trading. Third, attention becomes more informative
within collections when it is measured as sustained engagement rather than as a one-period impulse:
when attention is constructed as a smoothed buildup over multiple windows, the within-collection
association strengthens and becomes economically meaningful. Across alternative timing windows,
sample perturbations, and model variants, the central message remains intact: discourse measures
contain incremental information about valuation in this market, while the strength and direction of
short-run within-collection relationships can vary with market activity and collection-specific
context.

Several limitations qualify interpretation. The sentiment indicators are lexicon-based and therefore
cannot fully capture community-specific language, sarcasm, or shifting meanings across subcultures.
The bin-based timing approach is useful for defining lags in the presence of incomplete timestamps,
but it can smooth high-frequency dynamics and cannot perfectly map discourse to calendar time. The
analysis also focuses on Reddit and on a set of large collections, so results may not generalize to
other communication venues or to smaller and less liquid communities. Finally, due to resource
constraints, the paper does not deploy large language models to produce richer semantic
interpretations of discourse, which could reduce measurement error by separating distinct narrative
types such as marketing, coordination, conflict, and information production.

These results suggest several natural directions for future research. One priority is improved text
measurement: LLM-based classification, embeddings, and topic models could identify narrative content
with greater precision and allow sentiment to be measured in a context-aware way. A second priority
is timing and mechanism: with higher-resolution timestamps and explicit event-study designs, future
work could better distinguish reactive from anticipatory discourse and clarify when discussion
behaves more like demand pressure, liquidity coordination, or belief updating. A third priority is
platform integration: combining signals across Reddit, Discord, X, and marketplace metadata would
allow researchers to compare which channels contain the most pricing-relevant information and how
information propagates across communities. Finally, given the evidence that short-run relationships
differ in higher-activity states, regime-dependent models that explicitly condition on liquidity,
market phases, and collection norms may offer a more complete description of when and how social
information becomes priced in markets for digital cultural goods.

\subsection*{References}

\begingroup
\setlength{\parindent}{0pt}
\setlength{\parskip}{0.35\baselineskip}
\hangindent=1.5em
\hangafter=1
\Urlmuskip=0mu plus 1mu\relax
\makeatletter
\g@addto@macro\UrlBreaks{\do\/\do-\do\_\do\.\do\:\do\?\do\&\do\=\do\#}
\makeatother
\sloppy

Antweiler, W., \& Frank, M. Z. (2004). Is all that talk just noise? The information content of Internet stock message boards. \textit{Journal of Finance}, 59(3), 1259–1294. \url{https://doi.org/10.1111/j.1540-6261.2004.00662.x}

Ashenfelter, O., \& Graddy, K. (2003). Auctions and the price of art. \textit{Journal of Economic Literature}, 41(3), 763–787. \url{https://doi.org/10.1257/002205103322436188}

Bagwell, L. S., \& Bernheim, B. D. (1996). Veblen effects in a theory of conspicuous consumption. \textit{American Economic Review}, 86(3), 349–373. \url{https://doi.org/10.2307/2118201}

Barber, B. M., \& Odean, T. (2008). All that glitters: The effect of attention and news on the buying behavior of individual and institutional investors. \textit{Review of Financial Studies}, 21(2), 785–818. \url{https://doi.org/10.1093/rfs/hhm079}

Bikhchandani, S., Hirshleifer, D., \& Welch, I. (1992). A theory of fads, fashion, custom, and cultural change in informational cascades. \textit{Journal of Political Economy}, 100(5), 992–1026. \url{https://doi.org/10.1086/261849}

Chanel, O., Gérard-Varet, L.-A., \& Ginsburgh, V. (1996). The relevance of hedonic price indices: The case of paintings. \textit{Journal of Cultural Economics}, 20(1), 1–24. \url{https://doi.org/10.1007/s10824-005-1024-3}

Chen, Y., Ye, Y., \& Zeng, W. (2023). The rich, the poor, and the ugly: An aesthetic-perspective assessment of NFT values. In \textit{Proceedings of the 16th International Symposium on Visual Information Communication and Interaction (VINCI 2023)}. ACM. \url{https://doi.org/10.1145/3615522.3615545}

Da, Z., Engelberg, J., \& Gao, P. (2011). In search of attention. \textit{Journal of Finance}, 66(5), 1461–1499. \url{https://doi.org/10.1111/j.1540-6261.2011.01679.x}

Dowling, M. (2022a). Fertile LAND: Pricing non-fungible tokens. \textit{Finance Research Letters}, 44, 102096. \url{https://doi.org/10.1016/j.frl.2021.102096}

Dowling, M. (2022b). Is non-fungible token pricing driven by cryptocurrencies? \textit{Finance Research Letters}, 44, 102097. \url{https://doi.org/10.1016/j.frl.2021.102097}

Heffetz, O. (2011). A test of conspicuous consumption: Visibility and income elasticities. \textit{Review of Economics and Statistics}, 93(4), 1101–1117. \url{https://doi.org/10.1162/REST_a_00116}

Horky, F., Rachel, C., \& Fidrmuc, J. (2022). Price determinants of non-fungible tokens in the digital art market. \textit{Finance Research Letters}, 48, 103007. \url{https://doi.org/10.1016/j.frl.2022.103007}

Hubbard, T. P., \& LaFave, D. R. (2025). Price differentials and skin tone in digital art. \textit{Journal of Cultural Economics} (online first). \url{https://doi.org/10.1007/s10824-025-09570-0}

Mei, J., \& Moses, M. (2002). Art as an investment and the underperformance of masterpieces. \textit{American Economic Review}, 92(5), 1656–1668. \url{https://doi.org/10.1257/000282802762024719}

Nadini, M., Alessandretti, L., Di Giacinto, F., Martino, M., Aiello, L. M., \& Baronchelli, A. (2021). Mapping the NFT revolution: Market trends, trade networks, and visual features. \textit{Scientific Reports}, 11, 20902. \url{https://doi.org/10.1038/s41598-021-00053-8}

Rosen, S. (1974). Hedonic prices and implicit markets: Product differentiation in pure competition. \textit{Journal of Political Economy}, 82(1), 34–55. \url{https://doi.org/10.1086/260169}

Tariq, S. (2025). Pixels to prices: Visual traits, market cycles, and the economics of NFT valuation. \textit{arXiv} (arXiv:2509.24879). \url{https://doi.org/10.48550/arXiv.2509.24879}

Tetlock, P. C. (2007). Giving content to investor sentiment: The role of media in the stock market. \textit{Journal of Finance}, 62(3), 1139–1168. \url{https://doi.org/10.1111/j.1540-6261.2007.01232.x}

\endgroup

\end{document}